\documentstyle[preprint,aps]{revtex}

\begin{document}

\title{Isotope yields from central \\ $^{112,124}$Sn+$^{112,124}$Sn collisions, dynamical emission?}

\author{T.X. Liu, X.D. Liu, M.J. van Goethem, W.G. Lynch, R. Shomin, W.P. Tan, M.B. Tsang,
G. Verde, A. Wagner\footnote{
Present address: Institut f\"{u}r Kern- und Hadronenphysik,
Forschungszentrum Rossendorf, D-01314 Dresden, Germany}, H.F. Xi\footnote{
Present address: Benton Associate, Toronto, Ontario, Canada.}, H.S. Xu}

\author{\small National Superconducting Cyclotron Laboratory and
Department of Physics and Astronomy,
Michigan State University, East Lansing, MI 48824, USA}

\author{M. Colonna, M. Di Toro}

\author{\small Laboratorio Nazionale del Sud, Via S. Sofia 44, I-95123
Catania, Italy and University of Catania, I-95123 Catania, Italy}

\author{M. Zielinska-Pfabe}

\author{\small Department of Physics, Smith College, Northampton, MA 01063
USA}

\author{H. H. Wolter}

\author{\small Ludwig-Maximilians-Universit\"{a}t, M\"{u}nchen, Germany}

\author{L. Beaulieu, B. Davin, Y. Larochelle, T. Lefort, R. T. de Souza,
R. Yanez, and V. E. Viola}

\author{\small Department of Chemistry and IUCF, Indiana University,
Bloomington, IN 47405, USA}

\author{R.J. Charity, and L.G. Sobotka}

\author{\small Department of Chemistry, Washington University, St. Louis,
MO 63130, USA}

\date{\today}

\maketitle

\begin{abstract}
Isotopic yields for light particles and intermediate mass fragments have
been measured for $^{112}$Sn+$^{112}$Sn, $^{112}$Sn+$^{124}$Sn, $^{124}$Sn+$^{112}$Sn and $^{124}$Sn+$^{124}$Sn
central collisions at E/A=50 MeV and compared with predictions of stochastic
mean field calculations. These calculations predict a sensitivity of the
isotopic distributions to the density dependence of the asymmetry term of
the nuclear equation of state. However, the secondary decay of the excited
fragments modifies significantly the primary isotopic distributions and these
modifications are rather sensitive to theoretical uncertainties in the
excitation energies of the hot fragments. The predicted final isotope distributions are narrower than the experimental
data and the sensitivity of the predicted yields to the density dependence
of the asymmetry term is reduced.
\end{abstract}

\pacs{---}

\newpage

\noindent
{\bf Introduction}

The density dependence of the asymmetry term in the nuclear equation of
state (EOS) is an important but poorly constrained property of nuclear
matter [1-3]. Nuclear structure data provide few constraints on the density
dependence of the asymmetry term [3]. On the other hand, the asymmetry term
and its density dependence govern the density, radius and proton fraction of
neutron stars [2] and provide strong motivations for theoretical and
experimental investigations of these issues. Recently, a number of
calculations have tried to identify experimental observables, which can
provide constraints on the density dependence of the asymmetry term [4-7].

In this paper, we focus on sensitivities that have been predicted for
observables in energetic central nucleus collisions [4-7].  At incident
energies greater than about 35A MeV, the central density in such collisions
initially increases as the projectile and target nuclei overlap and then
decreases as the system collectively expands and eventually multi-fragments.
Previous investigations have shown that excited systems produced in such
collisions undergo bulk multifragmentation characterized by a short breakup
time scale 100 fm/c [8-11] and final states containing more than four
fragments of charge Z $>$ 2 [12,13].

For heavy systems in which the neutron density exceeds the proton density,
the asymmetry term is repulsive for neutrons and attractive for protons. The
asymmetry term therefore enhances the dynamical emission of neutrons
relative to protons in such collisions; the degree of enhancement reflects
the magnitude of the asymmetry term and its density dependence [4-7]. The
difference between neutron and proton emission rates in such collisions can
either be probed by direct measurements of pre-equilibirum neutron and
proton spectra or by examining the isotopic composition of the bound
fragments that remain after emission [4-7]. In this paper, we will
concentrate on the fragment observables.

Fragment observables for these reactions have been described successfully
via either statistical [14-17] or dynamical [18,19] models. To investigate
the dependence of such observables on the density dependence of the
asymmetry term, it is necessary to calculate the relative emission of
neutrons and protons and assess the change in the isotopic composition in
the prefragment during the expansion stage prior to the multifragment
breakup [4]. Within the context of a hybrid model, this pre-equilibrium
emission was calculated by ref. [20] using a Boltzmann-Uehling-Uhlenbeck
(BUU) equation model of ref. [4] and the subsequent fragmentation was
explored using the equilibrium Statistical Multifragmentation Model (SMM) of
ref. [21]. In a second approach [22], the expansion and multifragmentation
of the system was calculated by a statistical rate equation approach for
surface emission called the Expanding Emitting Source (EES) model of ref.
[14]. Both of these calculations predicted that the final isotopic
composition of observed fragments should be sensitive to the density
dependence of the asymmetry term.

The predicted sensitivities for the surface emission in the EES calculations
and for the bulk emission in the hybrid BUU-SMM calculations, however, are
completely opposite [20,22]. For calculations using an asymmetry term with
softer density dependence, the EES approach predicts preferential surface
emission of more neutron rich fragments while the BUU-SMM approach predicts
preferential bulk emission of more isospin symmetric fragments. These
differences stem from different model assumptions in these two approaches
about the density distribution of the system at the time of fragment
production. In the EES approach, the fragments are emitted at normal density
along with the protons and neutrons from a residue, which is at subnuclear
density. In the BUU-SMM approach, the fragments originate from the bulk
disintegration of the residue itself.

In this paper, we investigate how the isospin transport and dynamics is
related to the asymmetry term within the dynamical stochastic mean field
theory (SMF) approach wherein the evolution of the density and nuclear mean
field is calculated self consistently [7,23]. In previous publications,
stochastic mean field theory predictions for the influence of the asymmetry
term on fragment production, collective flow, incomplete fusion and binary
collisions have been reported [24-27]. Here, we compare this model to
isotopically resolved multifragmentation data measured in central
$^{124}$Sn+$^{124}$Sn and $^{112}$Sn+$^{112}$Sn reactions at 50A MeV. These experimental
multifragmentation data are presented in the next section. This is followed
by a description of the SMF approach, which provides predictions for the
dynamical production of highly excited fragments, and of the decay of these
excited fragments via the MSU statistical decay code [28,29]. The paper
concludes with a discussion of the comparison between data and theory and
the issues that remain for future investigations.

\vspace{.5 cm}

\noindent
{\bf The experiment}

Central $^{112}$Sn+$^{112}$Sn, $^{112}$Sn+$^{124}$Sn, $^{124}$Sn+$^{112}$Sn and $^{124}$Sn+$^{124}$Sn collisions
were measured at the National Superconducting Cyclotron Laboratory at
Michigan State University, using 5 mg/cm$^2$ $^{112}$Sn and $^{124}$Sn targets and 50 MeV
per nucleon $^{112}$Sn and $^{124}$Sn beams. Isotopically resolved light particles and
intermediate mass fragments with 3$\leq$Z$\leq$8 were measured with the Large Area
Silicon Strip detector Array (LASSA) [30,31], an array consisting of 9
telescopes, each comprised of one 65 $\mu$m and one 500 $\mu$m Si strip detector,
followed by four 6-cm thick CsI(Tl) detectors. The 50mm x 50mm lateral
dimensions of each LASSA telescope are divided by the strips of the second
silicon detector into 256 (3x3 mm2) square pixels, providing an angular
resolution of about 0.43$^o$. The LASSA device was centered at a polar angle of
$\theta_{lab}$=32$^o$ with respect to the beam axis, providing coverage at
polar angles of 7$^o\leq\theta_{lab}\leq$58$^o$.
At other angles, charged particles were detected in 188 plastic
scintillator - CsI(Tl) phoswich detectors of the Miniball/Miniwall array
[32], which subtended polar angles 7$^o\leq\theta_{lab}\leq$160$^o$.
The Miniball/Miniwall array
provided isotopic resolution for H and He nuclei and elemental resolution
for intermediate mass fragments (IMF's) with 3$\leq$Z$\leq$20.

The total charged particle multiplicity detected in the two arrays was used
for impact parameter determination. Central collisions,
corresponding to a reduced impact parameter of $b/b_{max}\leq$0.2 [33],
were selected by a gate on the
top 4\% of the charged-particle multiplicity distribution. From cross
section measurements for such events, we estimate a value for
$b_{max}$=7.51 fm
by neglecting fluctuations and assuming that multiplicity decreases
monotonically with impact parameter. This would suggest that our impact
parameter selection corresponds to $b\leq$1.5 fm; however, multiplicity
fluctuations at fixed impact parameter may extend the included range of
impact parameters outward (up to $b\approx$3 fm).

In the following we present isotopically resolved differential
multiplicities for fragments emitted at center of mass angles of
70$^o\leq\theta_{CM}\leq$110$^o$.
At these angles, the coverage of the LASSA array is excellent; the only
losses occurred for fragments emitted at very low energies E/A$<$ 0.2 MeV in
the center of mass, corresponding to small laboratory angles of
$\theta_{lab}\approx$7$^o$.
This enabled accurate calculation of the detection efficiency for
70$^o\leq\theta_{CM}\leq$110$^o$; the
fragment spectra were fitted and the losses at low energies ($<$2\%) were
estimated and corrected. The data presented below have been corrected for
the losses below threshold, for inefficiencies in the solid angle coverage
and for multiple hits in the detector telescopes.

Figure 1 shows the measured average differential multiplicities of Li, Be,
B, C, N and O isotopes at 70$^o\leq\theta_{CM}\leq$110$^o$. In this figure, the solid squares and
circles show the $^{112}$Sn+$^{112}$Sn and $^{124}$Sn+$^{124}$Sn data, respectively. The
isotopic yields of $^{112}$Sn+$^{124}$Sn and $^{124}$Sn+$^{112}$Sn are essentially equal; they
have been averaged and are shown by the open diamonds. The x-axis, N-Z,
corresponds to the neutron excess of the nuclides. The peaks of the
distributions are located at isotopes with N=Z+1. The yields of the B, C
isotopes are offset by a factor of 10 and the yields of the N and O isotopes
are offset by a factor of 100 in the figure. As expected, more neutron rich
nuclides are produced by the neutron rich system, $^{124}$Sn+$^{124}$Sn,
while the opposite is true for emission from the proton-rich isotope system,
$^{112}$Sn+$^{112}$Sn. The experimental results indicate that the multiplicities of
IMF's are approximately 10-20\% larger for the $^{124}$Sn+$^{124}$Sn entrance channel
than for the $^{112}$Sn+$^{112}$Sn entrance channel, consistent with previous
observations at an incident energy of 40 MeV per nucleon [34].

In general, the drop from the peak toward more proton-rich isotopes is
rather steep especially for elements with even values of Z. The main
differences between the isotope yields for the four different systems are
observed in the tails of the isotope distributions, where it is greater than
a factor of 4 for $^{20}$O. Larger differences may be expected for even more
exotic isotopes, but the background in the present measurement due to
multiple hits in the LASSA telescopes does not allow for their accurate
determination.

Recently, it has been shown [35] that the isotopic yields for systems
produced at approximately the same excitation energy per nucleon or the same
temperature satisfy an isoscaling relationship. Specifically, the ratio
R$_{21}$(N, Z) = Y$_{2}$(N,Z)/Y$_{1}$(N, Z) constructed using
the isotope yields Y$_{i}$(N, Z) with neutron number N and proton
number Z from two different reactions denoted by the index i,
i=1-2, obeys a simple relationship [20,22,35,36]:
\begin{equation}
R_{21} (N,Z)=\,\;C\;e^{\alpha N+\beta Z}
\label{eq1}
\end{equation}
Here, C is an overall normalization factor and and are isoscaling
parameters. This parameterization is discussed in greater detail within the
isoscaling section below. If we adopt the convention that reaction 2 is more
neutron-rich than reaction 1, one expects $\alpha$ to be positive and $\beta$ to be
negative. We have adopted that convention here and have fitted the ratios of
the isotopic yields for these four systems to extract the corresponding
values for $\alpha$ and $\beta$. These values for and are given in Table 1.

Equation 1, with only three parameters, C, $\alpha$ and $\beta$ can be used to predict
the isotope yields of $^{112}$Sn+$^{112}$Sn as well as the mixed systems, $^{112}$Sn+$^{124}$Sn
or $^{124}$Sn+$^{112}$Sn, using the measured yields of one system, e.g. $^{124}$Sn+$^{124}$Sn.
To illustrate how well this parameterization relates the yields of these
four systems, we take the yields of the $^{124}$Sn+$^{124}$Sn system as a reference
and use those yields and the fitted values of $\alpha$ and $\beta$ to predict the yields for
the other four systems. The dash and dot-dashed lines in Fig. 1 are the
calculated yields for $^{112}$Sn+$^{112}$Sn and $^{112}$Sn+$^{124}$Sn, respectively. For this
limited range of asymmetry, these isoscaling parameters can be described by
a linear dependence on either the initial N/Z or the asymmetry parameter,
$\delta =\frac{N-Z}{N+Z}$
of the reactions [35]. The excellent agreement between the predicted yields
and the data suggests that such scaling law extrapolations may have useful
predictive power. For example, we expect that these scaling predictions can
be accurately extrapolated to other mass-symmetric systems of A=200-250 nucleons at the
same incident energy per nucleon but with very different isospin asymmetry.

\vspace{.5cm}

\noindent
{\bf Description of the Models}

Now we turn to the theoretical interpretation of these data. To study the
density dependence of the asymmetry term of the EOS, we adopt the viewpoint
of the Stochastic Mean Field (SMF) approach described in refs. [7,23]. In
this approach, the time evolution of the nuclear density is calculated by
taking into account both the average phase-space trajectory predicted by the
Boltzmann-Nordheim-Vlasov equation and the fluctuations of the individual
collision trajectories about this average that can be predicted by equations
of the Boltzmann-Langevin type.

The virtue of such a dynamical approach for the study of isotopic effects
lies in its self-consistency. The flow of neutrons and protons is calculated
under the influence of Coulomb and asymmetry terms, which reflect
self-consistently the motion of these nucleons. Several different density
dependences of the asymmetry term were explored from which two are selected
for presentation here. In both cases, the asymmetry term is approximated by
the form,
\begin{equation}
E_{sym} \left( \rho ,\delta \right) =S\left( \rho \right) \cdot \delta^2
\label{eq2}
\end{equation}
where for the asymmetry term with a stronger density dependence,
\begin{equation}
S\left( \rho \right) =a\cdot \left( \frac{\rho }{\rho _{0} } \right) ^{2/3}
+b\cdot \frac{2\left( \rho /\rho _{0} \right) ^{2} }{1+\left( \rho /\rho
_{0} \right)}.
\label{eq3}
\end{equation}
Here, $\rho$ is the physical density, $\rho_0$ the saturation density, a=13.4 MeV and b=19 MeV
[2,4,37]. In the following, we refer to this as the ``super stiff``
asymmetry term. For the asymmetry term with weaker density dependence,
\begin{equation}
S\left( \rho \right) =a\cdot \left( \frac{\rho }{\rho _{0} } \right) ^{2/3}
+$ 240.9$\rho $ -819.1$\rho ^{2}
\label{eq4}
\end{equation}
where a=12.7 MeV [25], we refer to this as the ``soft`` asymmetry
term. In Fig. 2, it can be seen that the two expressions are
nearly equal at saturation density but differ at densities that
are either much larger or smaller than $\rho_0$.

In addition to the asymmetry term, the SMF
calculations have a Skyrme type isoscalar mean field with a soft equation of
state for symmetric matter characterized by an incompressibility constant
K=201 MeV. The isoscalar mean field and the asymmetry term of these
equations of state are used for the construction of the initial ground state
and for the time evolution of the collision. The nucleon-nucleon collisions
by the residual interaction are calculated from an energy and angle
dependent parameterization of the free nucleon-nucleon interaction and the
isospin dependence of the Pauli-blocking is considered during these
collisions.

The calculation solves the transport equations by evolving test particles of
finite width. As mentioned above, we use a reduced number of test particles
(50 test particles per nucleon) in the present calculations to inject
numerical noise into the evolution. In test calculations, we alternatively
employed the fluctuation mechanism discussed in ref. [23], which involves
damping the numerical noise by utilizing a large number of test particles
and introducing explicitly physical noise according to thermal fluctuations.
It was checked that both methods lead to similar results. In contrast to the
BOB method of ref. [38], these methods of inserting fluctuations are well
suited to reactions at finite impact parameter because they do not
presuppose knowledge of the most unstable modes.

When the system expands and reaches the spinodal instability (after about
110-120 fm/c), the most unstable modes are amplified and initiate the
formation of fragments via spinodal decomposition. The evolution of the
system is continued after spinodal decomposition until freeze-out where the
number of dynamically produced fragments and their properties are finally
determined. The system is decomposed into fragments using essentially a
coalescence mechanism in coordinate space; specifically, fragments are
defined by regions of density in the final distributions that are above a
``cut-off`` density of 1/8$\rho_{0}$. By definition, the freeze-out time occurs when
the average calculated number of fragments saturates. This occurs about 260
fm/c after initial contact of projectile and target nuclei in the present
simulations. The excitation energy of the fragments is determined by
calculating the thermal excitation energy in a local density approximation.
The procedure is rather rough and will overestimate the excitation energy
particularly for light fragments.

Some of the important features of these calculations and of the prior
BUU-SMM and EES calculations can be understood simply by considering the
influence of the density dependence of the asymmetry term on the relative
emission rates of neutrons and protons. In all Sn+Sn collisions, the
symmetry energy in the liquid drop model is positive, i.e. repulsive. The
interaction contribution to the symmetry energy gives rise to a repulsive
contribution to the mean field potential for neutrons and an attractive
contribution to the mean field potential for protons. The mean field
potential for an asymmetry term with stronger density dependence is larger
at high density and smaller at low density than that for an asymmetry term
with weaker density dependence.

It is the low-density behavior that dominates the predictions for the
isoscaling parameter. As the system expands and eventually multi-fragments,
the prefragment remains at subnuclear densities for a long time while it is
emitting nucleons. The asymmetry term with weaker density dependence around
$\rho_0$ increases the difference between the neutron and proton emission rates
leading to a more symmetric prefragment than is produced by calculations
with the asymmetry term which has a stronger density dependence.

The SMF calculations are interesting because they are free, in principle, of
arbitrary assumptions about whether the fragments are formed at the surfaces
or from the bulk disintegration of the system. Comparisons between the
fragmentation dynamics for different asymmetry terms were reported in ref.
[39]. The trends of these calculations are consistent with the prefragment
isospin dependences discussed above. In particular, fragments produced in
calculations with an asymmetry term with strong density dependence tend to
be more neutron-rich than the fragments produced in calculations with an
asymmetry term with weak density dependence. In this respect, these
predictions are similar to the results of the BUU-SMM calculations of ref.
[20] and opposite to the results of the EES calculations of ref. [22].

The SMF fully dynamical formation of fragments should actually be
more sensitive than the hybrid BUU-SMM to the interplay of the
EOS, i.e. to the density dependence of the asymmetry term, with
the fragmentation process. In the hybrid BUU-SMM calculations, the
EOS is entering only in the ``pre-equilibrium`` nucleon emission
described above. In the SMF approach, we have not only this
isospin effect on fast particle emission but also the full
dynamics of the isospin fractionation mechanism during the cluster
formation. In a neutron-rich system, this leads to a different N/Z
``concentration`` in the liquid phase (the fragments are more
symmetric) and in the gas phase (nucleons and light ions, are more
neutron-rich) [16,40]. This effect is associated with the unstable
behavior of dilute asymmetric nuclear matter and so in this way we
have the chance of testing the EOS also at sub-saturation density.

Asymmetry terms with weaker density dependence around $\rho_{0}$ must show a faster
increase at low densities and so a larger isospin fractionation/distillation
during the fragment formation [7]. Therefore in a fully dynamical picture of
fragmentation events a ``soft`` behavior of the asymmetry term around
saturation density will enhance the formation of more symmetric fragments
for two converging reasons: i) A larger `` preequilibrium''  neutron
emission rate as discussed before; ii) A stronger isospin
fractionation/distillation during the bulk disintegration. Opposite effects
are of course predicted for a rapidly increasing (``stiff``) asymmetry term
around $\rho_{0}$. In this sense we can expect the SMF results to be more sensitive
to the isospin dependences of the EOS at sub-saturation density.

At freezeout, the fragments are highly excited. For simplicity, we
assume that the de-excitation of these fragments can be calculated
as if the fragments are isolated. For this de-excitation stage, we
have tabulated the data table states, spins, isospins and
branching ratios for nuclei with Z$\leq$15. Where experimental
information is complete, it is used. Alternatively, empirical
level density expressions are used for the discrete levels. These
discrete levels are matched to continuum level density expressions
as described in ref. [28]. The decay of primary fragments with
Z$\geq$15 are calculated, following ref. [28], using known
branching ratios, when available, and using the Hauser-Feshbach
formalism when the information is lacking. The decays of heavier
nuclei are calculated using the Gemini statistical decay code
[41].

While the SMF calculations predict the numbers and properties of
the hot fragments that are produced at breakup, the predictions
for the relative abundances of light clusters such as the isotopes
with Z=1-2 that are emitted before the system expands to
sub-nuclear density are not very realistic. This prevents a
precise modeling of complete events including their detection
efficiency and means that the impact parameter selection based on
multiplicity cannot be imposed straightforwardly on the calculated
events as on the data. This and the considerable numerical effort
it requires have persuaded us to limit our comparisons to
calculations composed of 600 events for each of the
$^{112}$Sn+$^{112}$Sn and $^{124}$Sn+$^{124}$Sn reactions at a
fixed impact parameter of b=2 fm. We note, however, that the
widths in the multiplicity distributions at fixed impact parameter
are large enough that a range of impact parameters may contribute
significantly to the experimental data. Future calculations will
be necessary to assess quantitatively the importance of this
impact parameter smearing.

\vspace{.5cm}

\noindent
{\bf Overall behavior predicted by the SMF calculations}

In Fig. 3, the solid circles and open squares in the left panel show the
measured elemental multiplicities for 2$\leq$Z$\leq$8 averaged over
70$^o\leq\theta_{CM}\leq$110$^o$
for $^{124}$Sn+$^{124}$Sn and $^{112}$Sn +$^{112}$Sn collisions,
respectively.
The right panels show
the corresponding measured multiplicities as a function of the fragment
mass. These averaged multiplicities were obtained by summing the isotopic
multiplicities for 2$\leq$Z$\leq$8.

The dashed lines denote the corresponding distributions of hot primary
fragments calculated by the SMF model using the super stiff EOS. Due to the
low total number of events, we averaged these calculations over a slightly
larger angular interval of 60$^o\leq\theta_{CM}\leq$120$^o$. The solid lines show the
multiplicities of cold fragments after secondary decay. The statistical
uncertainties in these calculations are shown in the figure as vertical
bars. The corresponding uncertainties in the data are smaller than the data
points. If the angular integration was performed over the entire solid
angle, the averaged calculated multiplicities are about 20-30\% larger. This
difference reflects an anisotropy in the calculated primary angular
distributions for the heavier fragments. In the present calculations,
however, we do not have the capability to accurately calculate the
modifications of the angular distribution due to secondary decay so we
presently cannot explore this issue more quantitatively. As we will show
later, this anisotropy has no impact on the shapes of the isotopic
distributions for Z=3-8.

In general, the calculated primary and secondary fragment multiplicities are
smaller than the measured values for the lighter fragments Z=3,4 and are
somewhat closer to the measured values for Z=6-8. The lighter fragments with
Z$<$4 are mainly produced in secondary decay stage of the theoretical
calculations; the primary yields of these light fragments are much smaller
relative to the final yields than are the values for the heavier fragments.
Because the fragment multiplicities and angular distributions depend on
impact parameter, the comparison shown in Fig. 3 may be sensitive to the
impact parameter ranges included in both calculation and data. Future
calculations over a wide range of impact parameters are needed to address
this issue. Concerning the greater discrepancy for Z=2-4 fragments, we have
already noted that the formation of light clusters in the dynamical stage
before breakup is not well described in BUU- and SMF-type simulations,
because the unique structural properties of these fragments are not therein
well treated. (Treatments of the emission of light clusters in coupled
transport equations for nucleons and light clusters can be found in refs.
[42,43] and in the framework of FMD [44] or AMD [45] simulations.) On the
other hand, there is a considerable emission of protons and neutrons during
this stage; the total emission and consequently the asymmetry of the
remaining source may still be realistic.

Now we turn to an examination of calculated isotopic yields. The upper left
panel of Fig. 4 shows 
the isotopes of carbon nuclei predicted by the SMF calculations
over the entire angular range for $^{124}$Sn+$^{124}$Sn (solid
line) and $^{112}$Sn+$^{112}$Sn (dashed line); the dotted-dashed
and dotted lines show the corresponding calculations over the
60$^o\leq\theta_{CM}\leq$120$^o$ gate. Not surprisingly, the more
neutron rich $^{124}$Sn+$^{124}$Sn system preferentially produces
the more neutron rich isotopes. The peak of the carbon primary
distribution for the $^{124}$Sn+$^{124}$Sn system occurs at about
$^{15}$C while the peak for $^{112}$Sn+$^{112}$Sn system occurs at
lower mass, i.e. somewhere between $^{13}$C and $^{14}$C. The
differences between the angle gated and total primary yields are
small, and these difference translate into negligible differences
in the shape of the isotopic yield distribution after secondary
decay; we therefore do not plot the gated data because the two
curves are indistinguishable when normalized to each other. As the
statistics make it difficult to perform comparisons to isotopic
yields with a 60$^o\leq\theta_{CM}\leq$120$^o$ gate imposed, the
remaining calculated multiplicities in the paper are integrated
over the entire solid angle.

After sequential decays, one obtains the secondary distributions shown in
the lower left panel. No longer is there a noticeable difference between the
peak locations (at $^{12}$C in both systems); instead, the main differences are
found in the shape of the distribution, which is higher in the neutron-rich
isotopes and lower in the neutron-deficient isotopes for the $^{124}$Sn+$^{124}$Sn
system than it is for the $^{112}$Sn+$^{112}$Sn system. Such trends are also
qualitatively observed in the experimental data shown for the $^{124}$Sn+$^{124}$Sn
system by the solid circles and for the $^{112}$Sn+$^{112}$Sn system by the open
squares in the lower left panel. However, the experimental distributions are
considerably wider and more neutron rich than the model predictions. This
trend is replicated in the elemental distributions for all of the other
measured elements.

Another way to quantify the differences in the isotope distributions is by
the asymmetry parameter $\delta$=(N-Z)/(N+Z). The average asymmetry $<\delta>$ of the
isotopic distribution for each element is shown as a function of Z in the
right panels of Fig. 4. Following the same convention as in the left panels,
the solid and dashed lines show the average asymmetries for $^{124}$Sn+$^{124}$Sn and
$^{112}$Sn+$^{112}$Sn collisions; the upper and lower panels present results for the
primary and secondary fragment distributions, respectively. The calculated
differences between the two systems are more pronounced prior to secondary
decay than afterwards. The asymmetries of the corresponding data, shown for the $^{124}$Sn+$^{124}$Sn
system by the solid circles and for the $^{112}$Sn+$^{112}$Sn system by the open
squares in the lower right panel, are larger and display a stronger
dependence on the asymmetry of the system than do the final calculated
fragment yields after secondary decay.

\vspace{.5cm}

\noindent
{\bf Isoscaling analyses}

A more sensitive way to compare isotopic distributions is to construct the
isotopic ratio R$_{21}$(N,Z)=Y$_{2}$(N,Z)/Y$_{1}$(N,Z)
from the isotope yields Y$_{i}$(N,Z) with neutron number
N and proton number Z from two different reactions.
As discussed in the experimental section, R$_{21}$(N,Z) obeys a simple
relationship
$R_{21}(N,Z)=\,\;C\;e^{\alpha N+\beta Z}$
where $C$ is an overall normalization factor and $\alpha$ and $\beta$ are isoscaling
parameters [20,22,35,36]. Such an isoscaling relationship can be obtained in
statistical theories for two systems that are at the same temperature when
they produce fragments. Binding energy factors common to the yields for the
fragments in each system are cancelled by the ratio when the temperatures
are equal, leaving terms related to the chemical potentials or separation
energies [22]. In grand canonical models of multifragmentation
$\alpha =\frac{\Delta \mu _{p} }{T}$
and
$\beta =\frac{\Delta \mu _{n} }{T}$,
for example, where n,p are the differences in the chemical potentials for
the neutrons and protons in the two systems and T is the temperature
[20,22,35,36]. In some calculations [36], the values for the isoscaling
parameters extracted from equilibrium multifragmentation models are
similar before and after sequential decays, an observation that has been
attributed to a partial cancellation of secondary decay effects [35, 36].

While isoscaling can be expected for many statistical processes
[20,22,35,36], the question of whether it can be expected for specific
dynamical calculations remains open. To investigate whether the SMF
dynamical model displays isoscaling, we construct the relative isotope
ratios, R$_{21}$,primary, using the primary fragments produced in $^{124}$Sn+$^{124}$Sn
collisions as reaction 2 (numerator) and in $^{112}$Sn+$^{112}$Sn collisions as
reaction 1 (denominator).

The results are shown in the upper panel of Fig. 5. The error bars
reflect the statistical uncertainties. The predicted isotope
ratios for these primary fragments depend very strongly on the
neutron number and follow trends that appear consistent with
isoscaling relationship defined by Eq. 1. The uncertainties are
large reflecting the low statistics of the simulations, but the
strong dependence on neutron number makes it possible to discern
apparent isoscaling trends nonetheless. The lines are best fits
using Eq. 1 resulting in C=0.96, $\alpha$=1.07 and $\beta$=-1.43.
These values for $\alpha$ are much larger than values observed in
the experiment. The lower panel provides the corresponding SMF
predictions for the ratios, R$_{21,final}$, of the yields of
particle stable nuclei after secondary decay. For comparison
purposes, the scale for the ordinates of the top and bottom panels
are chosen to be the same; this demonstrates graphically that the
trends of the final isotope ratios are much flatter and the
corresponding isoscaling parameters ($\alpha$=0.286 and
$\beta$=-.288) are much smaller. Clearly, the isoscaling
parameters predicted by dynamic SMF calculations are strongly
modified by secondary data. This trend is very different from
equilibrium statistical models for multifragmentation where the
isoscaling parameters have been predicted to be insensitive to
secondary decay [35,36].

The isoscaling behavior of the dynamically produced fragments arises not
from thermal physics but rather from some special characteristics of the SMF
primary distributions predicted for these reactions. We find, for example,
the SMF primary isotopic and isotonic distributions can be roughly described
by Gaussians, see Fig. 3 [46]. Isotopic distributions, for example, can be
described by

\begin{equation}
Y\left( N,Z\right) =f(Z)\exp \left[ -\frac{\left( N-\bar{N} \left( Z\right)
\right) ^{2} }{2\sigma _{Z}^{2} } \right]
\label{eq5}
\end{equation}
where
$\bar{N} \left( Z\right)$
is the centroid of the distribution and
$\sigma _{Z}^{2}$
describes the width of the distribution for each for each element of charge
Z. This leads to an exponential behavior of the ratio R$_{21}$, since, neglecting
quadratic terms in $\bar{N}$,
\begin{equation}
\ln R_{21} =\frac{1}{\sigma _{Z}^{2} } \left[ \bar{N} \left( Z\right) _{2} -%
\bar{N} \left( Z\right) _{1} \right] N
\label{eq6}
\end{equation}
Note Eq. 6 requires the values for $\sigma _{Z}^{2}$
to be approximately the same for both reactions. We have observed this to be
the case for our SMF calculations of Sn+Sn collisions (to within the
statistical accuracy $\sim$10\%). For the ratios for every element, to be
optimally described by the same parameter, the ratio
$\left[ \bar{N} \left( Z\right) _{2} -\bar{N} \left( Z\right) _{1} \right]
/\sigma _{Z}$
must be independent of Z. The statistics of the calculation do not allow a
detailed test of this assumption, but it does appear that this ratio
increases somewhat with Z, as Fig. 5 confirms. The primary distributions
therefore do not respect the isoscaling relationship as well the data do.

At variance with the statistical fragmentation models, the
secondary decays substantially modify the isoscaling parameter.
The width $\sigma _{Z}^{2}$ decreases due to secondary decay and
the difference $\left[ \bar{N} _{2} -\bar{N} _{1} \right]$
likewise decreases fractionally, but by a larger amount. Moreover,
the final shape is no longer Gaussian, but due to secondary decay,
it reflects the binding energy as a function of neutron excess
more strongly (see Fig. 4). These changes combine to decrease the
isoscaling parameter as shown in the lower panel of Fig. 5.

\vspace{.5cm}

\noindent
{\bf Sensitivity of the SMF calculations to the asymmetry term}

The density dependence of the asymmetry term has a significant influence on
the relative emission rates of the neutrons and protons and, consequently,
on the isospin asymmetry of the hot fragments prior to secondary decay. As
discussed previously, an asymmetry term with weaker density dependence tends
to remain more important at lower densities, driving the fragments closer to
isospin symmetry, than does an asymmetry term with stronger density
dependence. Consistent with this general consideration, the calculated
primary isotope distributions in $^{124}$Sn+$^{124}$Sn collisions, shown in Fig. 6 for
carbon (upper left panel) and oxygen (upper right panel), are more
neutron-rich for the super stiff asymmetry term (solid line) than they are
for the soft asymmetry term (dashed line). A similar trend is also predicted
for the $^{112}$Sn+$^{112}$Sn system, but is not shown in the interest of brevity.

A similar trend is observed in the corresponding final distributions that
are obtained after secondary decay and shown in the middle panel with the
same convention for the solid and dashed lines as in the upper panel. Both
secondary distributions calculated for super stiff and soft asymmetry terms,
however, are significantly narrower and more proton-rich than the
experimental distributions shown by the closed circles in the figure. (The
lower panels, which display corresponding calculations when the excitation
energy is reduced by 50\%, will be discussed in the next section of this
paper.) Similar trends are also observed for the $^{112}$Sn+$^{112}$Sn and for
the other elements with 3$\leq$Z$\leq$8, though we do not for brevity's sake
show those results.

In Fig. 7, we present the related dependence of the SMF
predictions for the isotope ratios R$_{21}$ upon the density
dependence of the asymmetry term. We take advantage of the fact
that the results in the Fig. 5 can be compactly displayed by the
scaled function $S\left( N\right) =R_{21} \left( N,Z\right) \cdot
e^{-\beta Z}$ which condenses the isotopic dependence for the
various elements onto a single line [22]. The left panel in Fig. 7
shows the results for the super stiff asymmetry term and the right
panel shows the results for the soft asymmetry term. In each
panel, the values for S(N) obtained from the primary distribution
are shown by the symbols clustered about the dashed lines, the
results obtained from the secondary distribution are shown by the
symbols clustered about the dot-dashed lines and the results from
the data are shown by the solid lines in each panel to provide a
reference. Both the primary and secondary values for S(N) have
been fit by exponential functions to obtain corresponding values
for the scaling parameter and these values are given in the
figure.

Generally, the primary distributions for both
equations of state display a much stronger dependence on neutron number than
do the final isotopic distributions and the data. However, the influence on
the isoscaling parameter is statistically not very significant. Indeed, as
we pass from a `` stiff''  asymmetry term to a `` soft''  one, we do have a
stronger isospin fractionation/distillation, as already discussed before.
The centroid of the distribution,
$\bar{N} _{2}$, decreases (see Fig. 6) but also the width
$\sigma _{Z}^{2} $
decreases; the decrease in width, however, is of the order of 10\% and
comparable to its statistical uncertainty. The calculated final
distributions display a weak sensitivity to the density dependence of the
asymmetry term; the values for  ($\alpha$=0.286) obtained for the super stiff
asymmetry term are larger than the values for ($\alpha$=0.254) obtained for the soft
asymmetry term. The sensitivity to the asymmetry term is considerably less
than that reported for the EES model [22], and for the BUU-SMM hybrid
calculations [20]. Unlike these latter two calculations, both super stiff
and soft asymmetry terms yield $\alpha$ values that are significantly lower
than the value extrapolated from the data ($\alpha$=0.36). One should note, however,
that the excitation energies of these latter calculations could be more
freely varied to achieve better agreement with the experimental observations.

\vspace{.5cm}

\noindent
{\bf Discussion and summary}

The calculated final isotopic distributions for both asymmetry terms differ
from the measured ones in that they are narrower; more neutron deficient;
and show a weaker dependence on the isotopic asymmetry of the total system.
The last characteristic is reflected more clearly by the isoscaling
parameter than by direct examination of the isotopic distributions,
themselves. In these respects, the calculated results for the two different
asymmetry terms are more similar to each other than they are to the data. We
believe that it is probably premature at this stage to focus attention on
the sensitivity of the predicted final distributions to the asymmetry term.
Instead, let us concentrate upon what may be required to bring the final
isotopic distributions into greater concordance with the measurements.

The tendencies of the final isotopic distributions to be more
neutron deficient and to display a weaker dependence on the
isotopic asymmetry of the system are somewhat related. Both point
to difficulties the present calculations have in producing
neutron-rich isotopic distributions and indicate a surprising
sensitivity of the final results to the primary distributions.
That the final isotopic distributions are too neutron deficient
may result from the primary distributions being too neutron
deficient on the average, too narrow (i.e. $\sigma^{2} _{Z}$ is
too small) or that the secondary decay calculations predict too
much neutron emission after freezeout because the calculated
excitation energies are too high or the excitation energy
distributions are too narrow resulting in the loss of components
at low excitation energies that could decay to neutron-rich final
products.

We note that additional experimental measurements may help to
resolve these questions. The average isospin asymmetry of the
initial distributions is trivially related by charge and mass
conservation to the average isospin asymmetry of the nucleons
emitted during the SMF calculations before the freezeout (t=260
fm/c) chosen for these calculations. Complimentary measurements of
the yields and energy spectra of light particles can help to
determine whether these missing neutrons are carried away
primarily during pre-equilibrium emission during the
compression-expansion stage or during the later evaporative decay
of the hot fragments.

Previous authors have identified issues relevant to our calculations, which
may influence the asymmetries of the hot fragments at freezeout [43,47,48].
As discussed above, the present simulations underestimate the emission of
light clusters (d, t, $^{3}$He, $^{4}$He, $^{6}$Li, $^{7}$Li, etc.)
during the dynamical evolution prior to the freezeout.
Previous studies [43,47,48] have noted
that the neglect of the emission of $^{4}$He emission is particularly problematic
because it is abundant and because each $^{4}$He particle enhances the isospin
asymmetry of the remaining system by removing four nucleons without changing
the neutron excess. Indeed, it has been speculated that that $^{4}$He emission
may have an influence on the isospin asymmetry of the other clusters and
fragments that is of the same order of magnitude as the influence of the
mean field [43,47,48]. The issue needs additional theoretical attention.

Concerning neutron emission after freeze-out, we note that the number of
neutrons removed by secondary decay depends primarily on the fragment
excitation energies and the relative branching ratios for neutron and
charged particle emission. There are significant uncertainties in the
calculation of the excitation energies of the fragments, which are related
to the difficulty to establishing their precise ground state binding
energies. To explore the sensitivity of the results to the excitation
energy, we have reduced the excitation energy of each fragment by a
multiplicative factor f where 0.5$\leq$f$\leq$1 and recalculated the final fragment
isotopic distributions.

The solid and dashed lines in Fig. 6 for carbon fragments (bottom
left panel) and oxygen fragments (bottom right panel) show the
calculated final distributions for f=0.5 using super stiff and
soft asymmetry terms, respectively. Clearly, it is possible by
reducing the excitation energy to shift the isotope distribution
in the direction of the more neutron-rich isotopes, so as to make
the mean isospin asymmetry of the calculated final and measured
distributions to be the same. However, the widths of the
calculated final isotopic distributions will still be narrower
than the measured ones.

This discrepancy between the theoretical and experimental widths would be
reduced if the theoretical primary distributions were wider in their
excitation energy distributions or wider in their isotopic distributions or both. It
is interesting that the primary distributions of equilibrium statistical
model calculations that reproduce the experimental final distributions are
much wider in excitation energy and neutron number than those predicted by
the SMF calculations [21,36]. Future investigations will be needed to
address whether wider primary distributions in excitation energy or neutron
number can be attained in the SMF model by altering some of the underlying
model assumptions such as the manner in which they are defined at freezeout.

Increased widths may be achieved by performing calculations for a
range of impact parameters rather than the single impact parameter
b=2 fm presented here. We note that the inclusion of larger impact
parameter events may broaden the primary distributions at
mid-rapidity because it will require the inclusion of fragments
emitted from the neck joining projectile- and target-like
residues. In ref. [7], it was shown that SMF calculations predict
such `` neck''  fragments to be more neutron rich because the
isospin fractionation/distillation effect is somewhat reduced in
peripheral events, leaving more neutrons in the fragments, and
there can also be an overall neutron enhancement in the neck
region for such events due to the neutron skins of the projectile
and target [7]. Moreover, the excitation energies of the primary
fragments in more peripheral collisions are also somewhat reduced.
Experiments also suggest that neck fragments in very-peripheral
collisions are more neutron-rich [49], but there is little
detailed experimental information about the dependence at smaller
impact parameters. Combining more calculated peripheral events
with the calculated central collision events presented in this
paper may result in the broader isotopic distributions that appear
to be required by the data.

In summary, we have measured the isotope distributions of Z=2-8 particles
emitted in four different Sn+Sn reactions with different isospin asymmetry
and have calculated them with a dynamical model that includes fluctuations
that give rise to fragment production. The experimental data display a
strong dependence on the isospin asymmetry that can be accurately described
by an isoscaling parameterization. The theoretical calculations reproduce
the yields for the heavier fragments with Z=6-8, but underpredict the yields
of the lighter ones, which are not strongly produced as primary fragments.
The calculated final isotopic distributions display isoscaling, but the
calculated isotopic distributions are narrower, more neutron deficient; and
show a weaker dependence on the isotopic asymmetry of the system than do the
data. The density dependence of the asymmetry term of the EOS has an effect
on the calculated final isotopic distributions. The distributions calculated
using the asymmetry term with stronger density dependence are more
neutron-rich and are closer to the measured values. These trends are similar
to prior results obtained for a BUU-SMM hybrid model, but different from the
trends for evaporated fragments predicted by EES rate equation calculations.
The present level of agreement between theory and experiment precludes
definitive statements about the density dependence of the asymmetry term of
the EOS, but it does reveal that that final distributions are surprisingly
sensitive to the widths predicted by the SMF model for the primary fragment
isotope and excitation energy distributions. A number of theoretical issues,
such as the pre-equilibrium emission of bound clusters, the calculations of
fragment excitation energies, the way fragments are defined at freezeout,
and the impact parameter range modeled by the calculation may influence the
calculated results. Additional theoretical work is required to explore these
issues and to determine the role they may play in the resolution of these
discrepancies. Complementary measurements of the isospin asymmetry of light
cluster emission prior to the multifragment breakup can provide information
relevant to the resolution of these issues.

This work is supported by the National Science Foundation under Grant Nos.
PHY-95-28844 and PHY-0070818, by the U.S. Department of Energy under
contract Nos. DE-FG02-92ER-40714 and DE-FG02-87ER-40316 and by the INFN of
Italy. Some of the authors benefited from the scientific environment of a
workshop at ECT*, Trento where this work was extensively discussed.

\newpage

{\bf REFERENCES}

\noindent
1. I. Bombaci, ``Equation of State for Dense Isospin Asymmetric Nuclear
Matter for Astrophysical Applications``, ``Isospin Physics in Heavy-IonCollisions at Intermediate Energies``, Eds. Bao-An Li and W. Udo Schroeder,
NOVA Science Publishers, Inc. (New York), p. 35 (2001).

\noindent
2. J.M. Lattimer and M. Prakash, Ap. J., 550, 426 (2001) and refs. therein.

\noindent
3. B. Alex Brown, Phys. Rev. Lett. 85, 5296 (2001).

\noindent
4. Bao-An Li, Phys. Rev. Lett. 85, 4221 (2000).

\noindent
5. L. Scalone, M. Colonna and M. Di Toro, Physics Letters B 461, 9 (1999).

\noindent
6. V. Baran, M. Colonna, M. Di Toro, and A.B. Larionov, Nucl. Phys. A 632,
287 (1998).

\noindent
7. V. Baran, M. Colonna, M. Di Toro, V. Greco, M. Zielinska-Pfab\'{e} and
H.H. Wolter, Nucl. Phys. A 703, 603 (2002).

\noindent
8. M. D'Agostino, A.S. Botvina, P.M. Milazzo, M. Bruno, G.J. Kunde, D.R.
Bowman , L. Celano, N. Colonna, J.D. Dinius, A. Ferrero, M.L. Fiandri, C.K.
Gelbke, T. Glasmacher, F. Gramegna, D.O. Handzy, D. Horn, W.C. Hsi, M.
Huang, I. Iori,  M.A. Lisa, W.G. Lynch, L. Manduci, G.V. Margagliotti, P.F.
Mastinu, I.N. Mishustin , C.P. Montoya, A. Moroni, G.F. Peaslee , F.
Petruzzelli, L. Phair, R. Rui, C. Schwarz, M.B. Tsang, G. Vannini, and C.
Williams, Phys. Lett. B 371, 175 (1996).

\noindent
9. D.R. Bowman, G.F. Peaslee, N. Carlin R.T. de Souza, C.K. Gelbke, W.G.
Gong, Y.D. Kim, M.A. Lisa, W.G. Lynch, L. Phair, M.B. Tsang, C. Williams, N.
Colonna, K. Hanold, M.A. McMahan, G.J. Wozniak, and L.G. Moretto, Phys. Rev.
Lett. 70,  3534 (1993).

\noindent
10. E. Cornell, T.M. Hamilton, D. Fox, Y. Lou, R.T. de Souza, M.J. Huang,
W.C. Hsi, C. Schwarz, C. Williams, D.R. Bowman, J. Dinius, C.K. Gelbke, T.
Glasmacher, D.O. Handzy, M.A. Lisa, W.G. Lynch, G.F. Peaslee, L. Phair, M.B.
Tsang, G. VanBuren, R.J. Charity, L.G. Sobotka, and W.A. Friedman, Phys.
Rev. Lett. 75, 1475 (1995).

\noindent
11. R. Popescu, T. Glasmacher, J.D. Dinius, S.J. Gaff, C.K. Gelbke, D.O.
Handzy, M.J. Huang, G.J. Kunde, W.G. Lynch, L. Martin, C.P. Montoya, M.B.
Tsang, N. Colonna, L. Celano, G. Tagliente, G.V. Margagliotti, P.M. Milazzo,
R. Rui, G. Vannini, M. Bruno, M. D'Agostino, M.L. Fiandri, F. Gramegna, A.
Ferrero, I. Iori, A. Moroni, F. Petruzzelli, P.F. Mastinu, L. Phair, and K.
Tso, Phys. Rev. 58, 270 (1998).

\noindent
12. G.J. Kunde, S. J. Gaff, C. K. Gelbke, T. Glasmacher, M. J. Huang, R.
Lemmon, W. G. Lynch, L. Manduci, L. Martin, and M. B. Tsang W. A. Friedman
J. Dempsey, R. J. Charity, and L. G. Sobotka D. K. Agnihotri, B. Djerroud,
W. U. Schroder, W. Skulski, J. Toke, and K. Wyrozebski, Phys. Rev. Lett. 77,
2897 (1996).

\noindent
13. N. Marie, A. Chbihi, J.B. Natowitz, A. Le F\`{e}vre, S. Salou, J.P.
Wieleczko, L. Gingras, M. Assenard, G. Auger, Ch.O. Bacri, F. Bocage, B.
Borderie, R. Bougault, R. Brou, P. Buchet, J.L. Charvet, J. Cibor, J. Colin,
D. Cussol, R. Dayras, A. Demeyer, D. Dor\'{e}, D. Durand, P. Eudes, J.D.
Frankland, E. Galichet, E. Genouin-Duhamel, E. Gerlic, M. Germain, D.
Gourio, D. Guinet, K. Hagel, P. Lautesse, J.L. Laville, J.F. Lecolley, T.
Lefort, R. Legrain, N. Le Neindre, O. Lopez, M. Louvel, Z. Majka, A.M.
Maskay, L. Nalpas, A.D. Nguyen, M. Parlog, J. P\'{e}ter, E. Plagnol, A.
Rahmani, T. Reposeur, M.F. Rivet, E. Rosato, F. Saint-Laurent, J.C.
Steckmeyer, M. Stern, G. Tabacaru, B. Tamain, O. Tirel, E. Vient, C. Volant,
and R. Wada, Phys. Rev. C 58, 256 (1998).

\noindent
14.W.A. Friedman, Phys. Rev. C 42, 667 (1990).

\noindent
15. J.P. Bondorf, A.S. Botvina, A.S. Iljinov, l.N, Mishustin, and K.
Sneppen, Phys. Rep. 257, 133 (1995) and refs. therein.

\noindent
16. H. M\"{u}ller and B. D. Serot, Phys. Rev. C 52, 2072 (1995).

\noindent
17. D.H.E. Gross, Phys. Rep. 279, 119 (1997).

\noindent
18. B. Borderie, G. Tbcaru,, Ph. Chomaz, M. Colonna, A. Guarnera, M. P\^{a}%
rlog, M. F. Rivet, G. Auger, Ch. O. Bacri, N. Bellaize, R. Bougault, B.
Bouriquet, R. Brou, P. Buchet, A. Chbihi, J. Colin, A. Demeyer, E.
Galichet,, E. Gerlic, D. Guinet, S. Hudan, P. Lautesse, F. Lavaud, J. L.
Laville, J. F. Lecolley, C. Leduc, R. Legrain, N. Le Neindre, O. Lopez, M.
Louvel, A. M. Maskay, J. Normand, P. Pawowski, E. Rosato, F.Saint-Laurent,
J. C. Steckmeyer, B. Tamain, L. Tassan-Got, E. Vient, and J. P. Wieleczko,
Phys. Rev. Lett. 86, 3252 (2001).

\noindent
19. R. Wada, K. Hagel, J. Cibor, M. Gonin, Th. Keutgen, M. Murray, J. B.
Natowitz, A. Ono, J. C. Steckmeyer, A. Kerambrum, J. C. Ang\'{e}lique, A.
Auger, G. Bizard, R. Brou, C. Cabot, E. Crema, D. Cussol, D. Durand, Y. El
Masri, P. Eudes, Z. Y. He, S. C. Jeong, C.  Lebrun, J. P. Patry, A. P\'{e}%
ghaire, J. Peter, R. R\'{e}gimbart, E. Rosato, F. Saint-Laurent, B. Tamain,
and E. Vient,  Phys. Rev. C 62, 034601 (2000).

\noindent
20. W. P. Tan, B.-A. Li, R. Donangelo, C. K. Gelbke, M.-J. van Goethem, X.
D. Liu, W. G. Lynch, S. Souza, M. B. Tsang, G. Verde, A. Wagner, and H. S.
Xu, Phys. Rev. C 64, 051901(R) (2001).

\noindent
21. S. R. Souza, W. P. Tan, R. Donangelo, C. K. Gelbke, W. G. Lynch, and M.
B. Tsang, Phys. Rev. C 62, 064607 (2000).

\noindent
22. M. B. Tsang, W. A. Friedman, C. K. Gelbke, W. G. Lynch, G. Verde, and H.
S. Xu. Phys. Rev. Lett. 86, 5023 (2001).

\noindent
23. M. Colonna, M. Di Toro, A. Guarnera, S. Maccarone, M. Zielinska-Pfab\'{e}
and H.H. Wolter, Nucl. Phys. A 642, 449 (1998).

\noindent
24. V. Baran, M. Colonna, M. Di Toro, and A.B. Larionov, Nucl. Phys. A 632,
287 (1998).

\noindent
25. M. Colonna, M. Di Toro, G. Fabbri, and S. Maccarone, Phys. Rev. C 57,
1410 (1998).

\noindent
26. A.B. Larionov, A.S. Botvina, M. Colonna, M Di Toro, Nucl. Phys. A 658,
375 (1999).

\noindent
27. L. Scalone, M. Colonna and M. Di Toro, Physics Letters B, Vol. 461 (1-2)
(1999) pp. 9.

\noindent
28. H. M. Xu, W. G. Lynch, C. K. Gelbke, M. B. Tsang, D. J. Fields, M. R.
Maier, D. J. Morrissey, T. K. Nayak, J. Pochodzalla, D. G. Sarantites, L. G.
Sobotka, M. L. Halbert and D. C. Hensley, Phys. Rev. C 40, 186 (1989).

\noindent
29. W. Tan et al., in preparation.

\noindent
30. B. Davin, R.T. de Souza, R. Yanez, Y. Larochelle, R. Alfaro, H.S. Xu, A.
Alexander, K. Bastin, L. Beaulieu, J. Dorsett, G.Fleener, L. Gelovani, T.
Lefort, J. Poehhman, R.J. Charity, L.G. Sobotka, J. Elson, A. Wagner, T.X.
Liu, X.D. Liu, W.G. Lynch, L. Morris, R. Shomin, W.P. Tan, M.B. Tsang, G.
Verde, J. Yurkon, Nucl. Inst. Meth. 473, 302 (2001).

\noindent
31. A. Wagner, W.P. Tan, K. Chalut, R.J. Charity, B. Davin, Y. Larochelle,
M.D. Lennek, T.X. Liu, X.D. Liu, W.G. Lynch, A.M Ramos, R. Shomin, L.G.
Sobotka, R.T. de Souza, M.B. Tsang, G. Verde, H.S. Xu, Nucl. Inst. Meth.
456, 290 (2001).

\noindent
32. R.T. de Souza et al., Nucl. Instrum. Methods A 295, 109 (1990). D.W.
Stracener et al., Nucl. Instrum. Methods A 294, 485 (1990).

\noindent
33. L. Phair et al., Nucl. Phys. A 548, 489 (1992) and references therein.

\noindent
34. G. J. Kunde, et al., Phys. Rev. Lett. 77, 2897 (1996).

\noindent
35. H. S. Xu, M. B. Tsang, T. X. Liu, X. D. Liu, W. G. Lynch, W. P. Tan, A.
Vander Molen, G. Verde, A. Wagner, H. F. Xi, C. K. Gelbke, L. Beaulieu, B.
Davin, Y. Larochelle, T. Lefort, R. T. de Souza, R. Yanez, V. E. Viola, R.
J. Charity, and L. G. Sobotka, Phys. Rev. Lett. 85, 716 (2000).

\noindent
36. M.B. Tsang, C.K. Gelbke, X.D. Liu, W.G. Lynch, W.P. Tan, G. Verde, H.S.
Xu, W.A. Friedman, R. Donangelo, S.R. Souza, C.B. Das, S. Das Gupta, D.
Zhabinsky, Phys. Rev. C 64, 054615 (2001).

\noindent
37. M. Prakash, T.L. Ainsworth, and J.M. Lattimer, Phys. Rev. Lett. 61, 2518
(1988).

\noindent
38. A. Guarnera, M. Colonna, P. Chomaz, Phys. Lett. B 373, 373 (1996); A.
Guarnera, Thesis U. Caen (1996).

\noindent
39. M. Di Toro, V. Baran, M. Colonna, S. Maccarone, M. Zielinska-Pfabe, H.H.
Wolter, Nucl. Phys. A 681, 426C(2001).

\noindent
40. V. Baran, M. Colonna, M. Di Toro, and V. Greco, Phys. Rev. Lett. 86,
4492 (1001).

\noindent
41. R.J. Charity, M.A. McMahan, G.J. Wozniak, R.J. McDonald, L.G. Moretto,
D.G. Sarantites, L.G. Sobotka, G. Guarino, A. Pantaleo, L. Fiore, A. Gobbi,
and K.D. Hildenbrand, Nucl. Phys. A 483, 371 (1988).

\noindent
42. M.B. Tsang, P. Danielewicz, D.R. Bowman, N. Carlin, C.K. Gelbke, W.G.
Gong, Y.D. Kim, W.G. Lynch, L. Phair, R.T. de Souza, and F. Zhu, Phys. Lett.
B 297, 243 (1992).

\noindent
43. P. Danielewicz and G. F. Bertsch, Nucl. Phys. A 533, 712 (1991).

\noindent
44. H. Feldmeier, K. Bieler, and J. Schnack, Nucl. Phys. A 586, 493 (1995),

\noindent
45. A. Ono, H. Horiuchi, T. Maruyama, and A. Ohnishi, Phys. Rev. Lett. 68,
2898 (1992); A. Ono and H. Horiuchi, Phys. Rev. C53, 2958 (1996).

\noindent
46. One might imagine that the fragment isobaric distributions might display
a Gaussian dependence on asymmetry, i.e.
$Y\left( N,Z\right) =f(A)\exp \left[ -\frac{\left( \delta -\bar{\delta }
\right) ^{2} }{2\sigma ^{2} } \right] $
where $\delta$=(N-Z)/A and
$\bar{\delta } =\left\langle \delta \right\rangle $
but we lack the statistics in our simulations to test this hypothesis.

\noindent
47. L.G. Sobotka, J.F. Dempsey, and R.J. Charity, Phys. Rev. C 55, 2109
(1997)

\noindent
48. L. Shi and P. Danielewicz, Europhysics Lett. 51, 34 (2000).

\noindent
49. J. F. Dempsey et al., Phys. Rev, C 54, 1710 (1996) and refs. therein.

\begin{table}[htbp]
\begin{center}
\begin{tabular}{|c|c|c|c|}
Reaction 2 & Reaction 1 & $\alpha$ & $\beta$ \\ \hline
$^{112}$Sn+$^{124}$Sn & $^{112}$Sn+$^{112}$Sn & 0.18$\pm$0.01
& -0.19$\pm$0.01 \\ \hline
$^{124}$Sn+$^{124}$Sn & $^{112}$Sn+$^{112}$Sn & 0.36$\pm$0.02
& -0.39$\pm$0.01 \\
\end{tabular}
\vspace{1cm}
\caption{Values for $\alpha$ and $\beta$ for obtained from fitting the isotope
ratios R$_{21}$.}
\end{center}
\end{table}

\newpage

\begin{center} {\bf FIGURE CAPTIONS:} \end{center}

\noindent
Figure 1: Average differential multiplicities at 70$^o\leq\theta_{CM}\leq$110$^o$ for Li, Be, B, C, N
and O isotopes as a function of neutron excess (N-Z) of the isotope. The
solid circles (connected by solid lines to guide the eye) are the data for the
$^{124}$Sn+$^{124}$Sn system with N/Z=1.48. The solid squares are data for the lightest
system $^{112}$Sn+$^{112}$Sn with N/Z=1.24. The open diamonds are the averaged values
from the two mixed systems, $^{124}$Sn+$^{112}$Sn and $^{112}$Sn+$^{124}$Sn. The dashed and
dot-dashed lines are predictions from Eq. 1. See text for more details.

\noindent
Figure 2: The solid curve and dashed curves indicate the density
dependencies for the super stiff and soft asymmetry terms, respectively.

\noindent
Figure 3: Differential multiplicities for $^{124}$Sn+$^{124}$Sn collisions (upper
panel) and $^{112}$Sn+$^{112}$Sn collisions (lower panel) as a function of the
fragment charge (left panels) and the fragment mass (right panels). The
points are the data. The dashed and solid lines are the calculated primary
and final fragment differential multiplicities, respectively. Statistical
uncertainties are shown for the calculations; the corresponding
uncertainties in the data are smaller than the data points.

\noindent
Figure 4: Left panels: Calculated primary (upper panel), calculated final
(lower panel) and measured (lower panel) carbon isotopic yields for Sn+Sn
collisions. Right panels: Calculated primary (upper panel), calculated final
(lower panel) and measured (lower panel) mean isospin asymmetries as a
function of the fragment charge for Sn+Sn collisions. The lines and data
points are further explained in the text.

\noindent
Figure 5: Upper panel: R$_{21}$ values obtained from the ratios of the primary
isotopic distributions for $^{124}$Sn+$^{124}$Sn collisions divided by those for
$^{112}$Sn+$^{112}$Sn collisions. Lower panel: Corresponding R$_{21}$ values obtained from
the ratios of the final isotopic distributions. Each line in the two panels
corresponds to ratios for a given element. Elements with Z=2-8 (Z=1-8) are
represented from left to right in the upper (lower) panel. The lines are the
result of fitting R$_{21}$ with Eq. 1; the dependencies on neutron number for the
best fits are given in each panel.

\noindent
Figure 6: Upper panel: Dependence of the primary distributions for
carbon(left panel) and oxygen (right panel) upon the density
dependence of the asymmetry term. Middle panel:
Dependence of the final distributions for
carbon (left panel) and oxygen (right panel) upon the density dependence of
the asymmetry term. The data are also shown as the solid points. The various
lines in the figure are described in the text. The excitation energies for
the fragments are taken directly from the SMF calculations. Lower panel: The
data are the same as in the middle panels. The curves are the calculations
obtained when the excitation energies of the primary fragments are reduced
by a factor of two.

\noindent Figure 7: Dependence of the scaled function S(N) on the
density dependence of the asymmetry term. The left panel provides
a comparison between values for S(N) computed from the data (solid
line) and the calculated primary (points about dashed line) and
final (points about dashed-dotted line) distributions obtained for
the super stiff asymmetry term. The right panel provides a
comparison between values for S(N) computed from the data and the
calculated primary and final distributions obtained for the soft
asymmetry term.

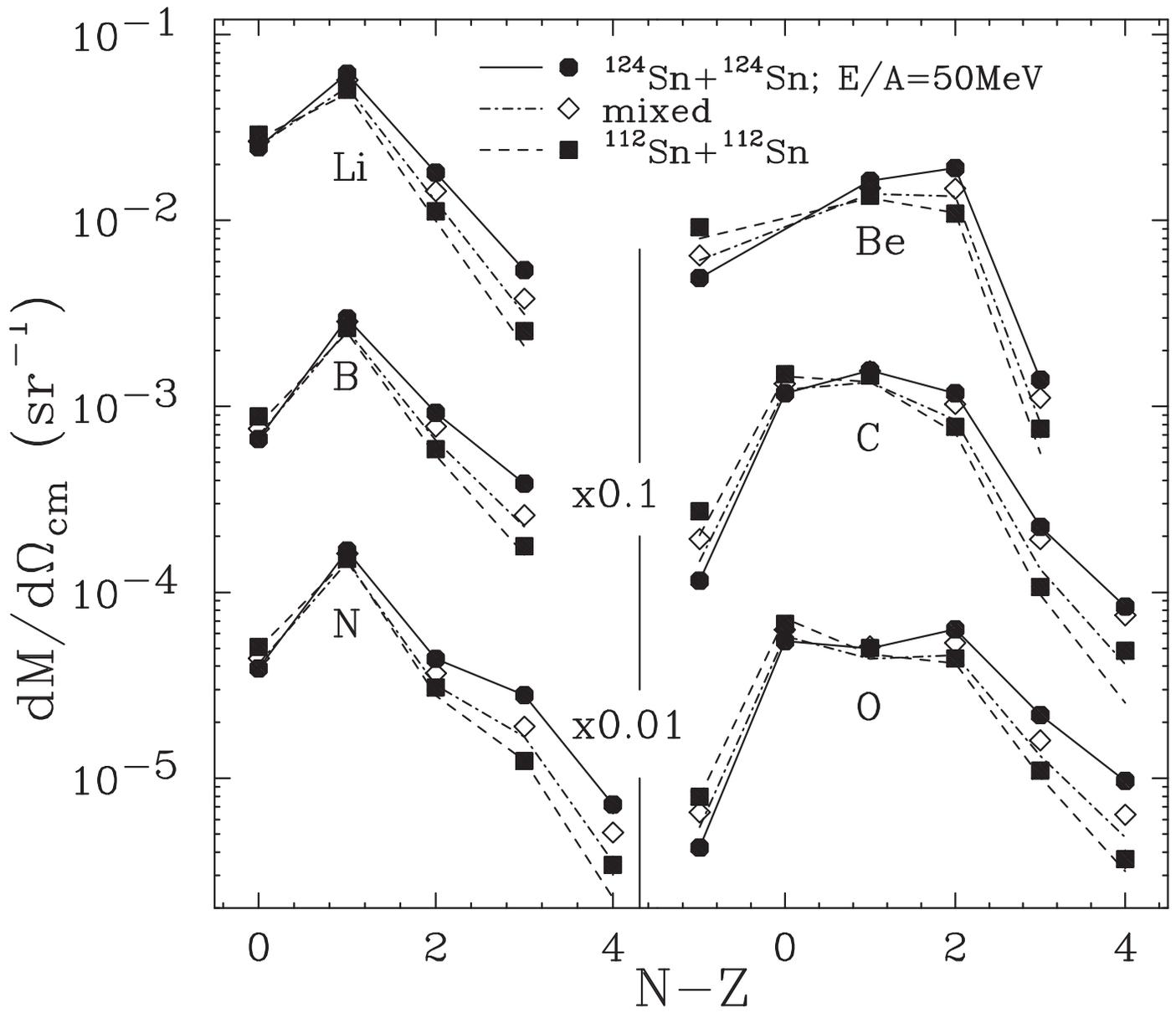
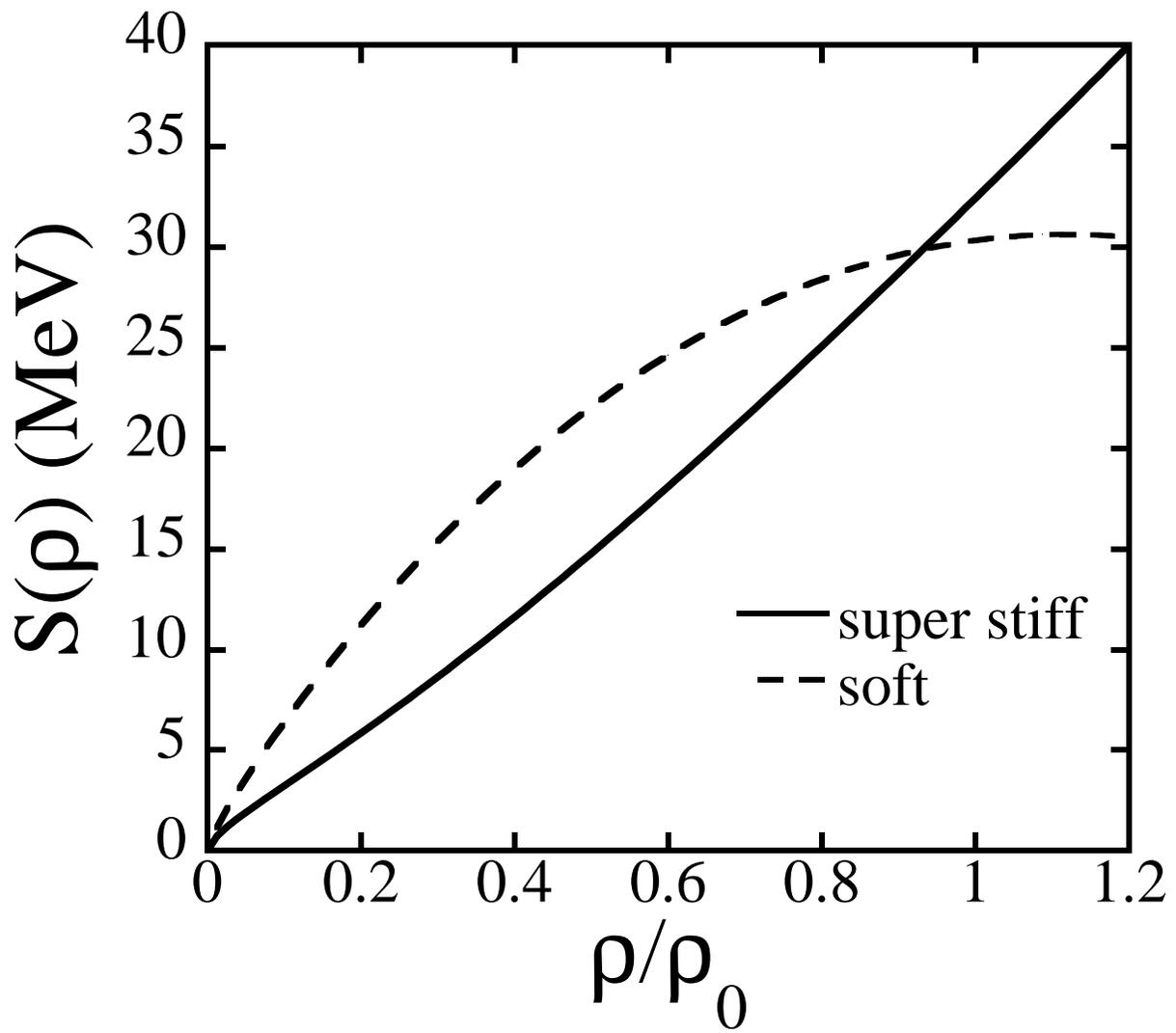
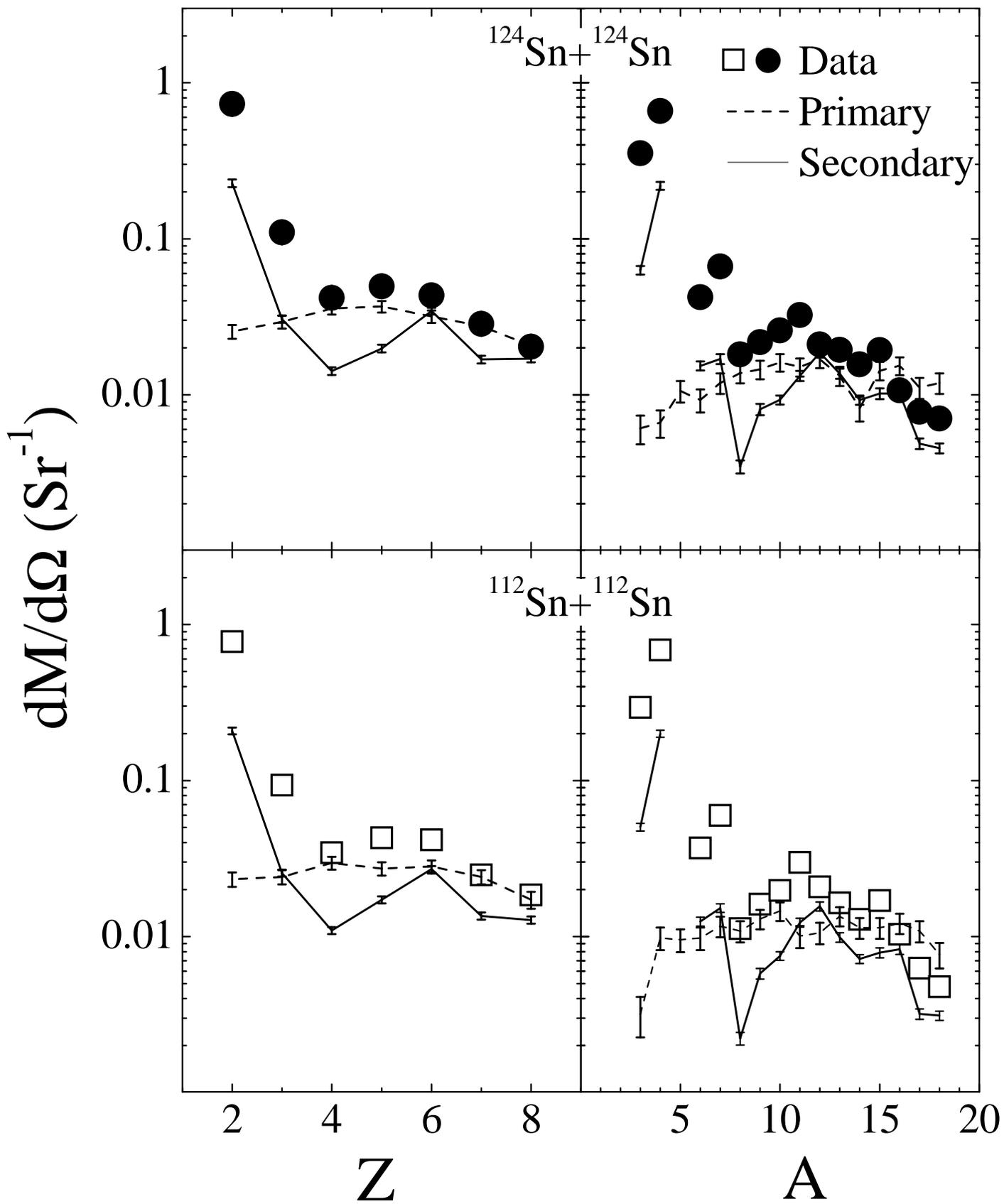
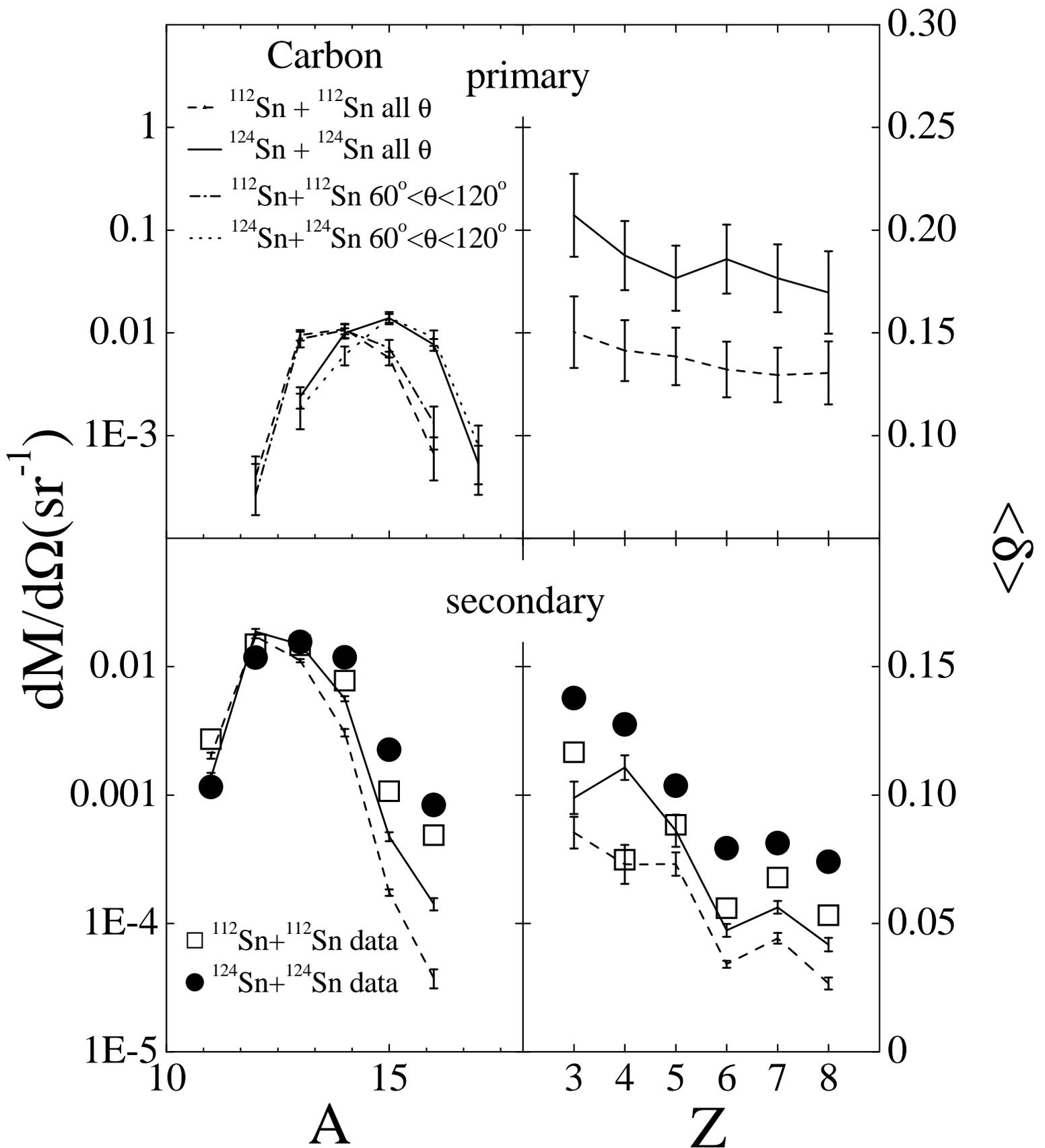
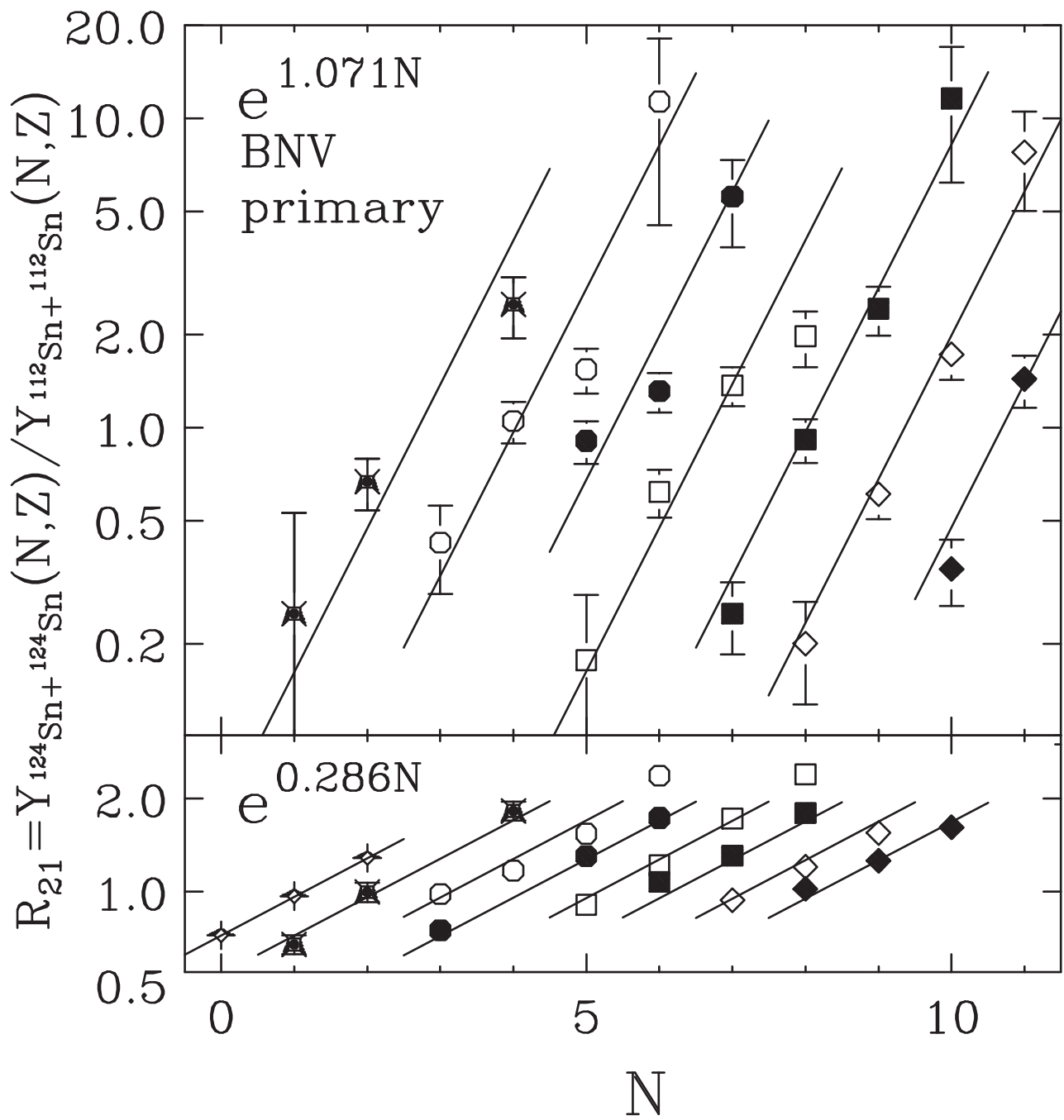
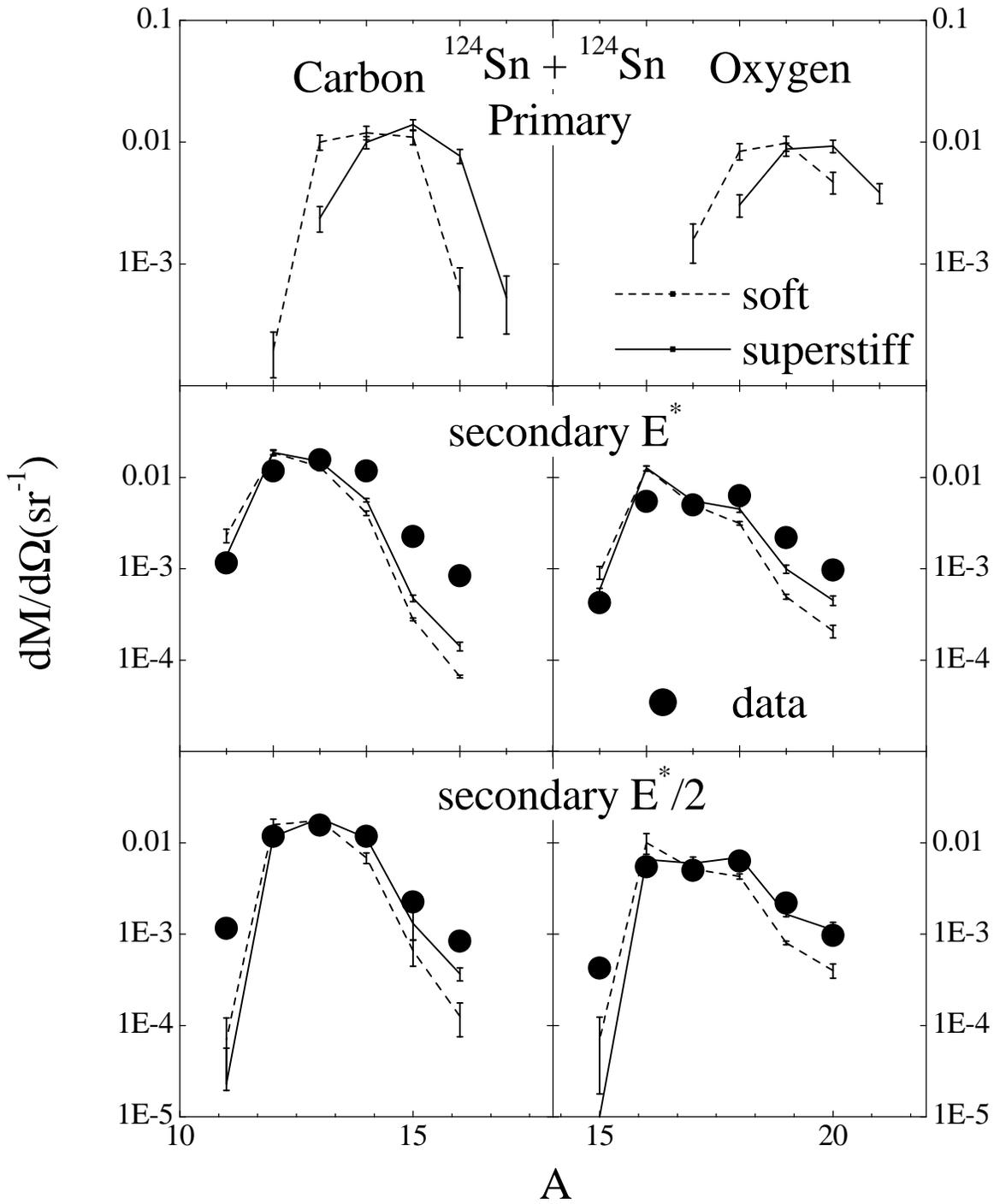
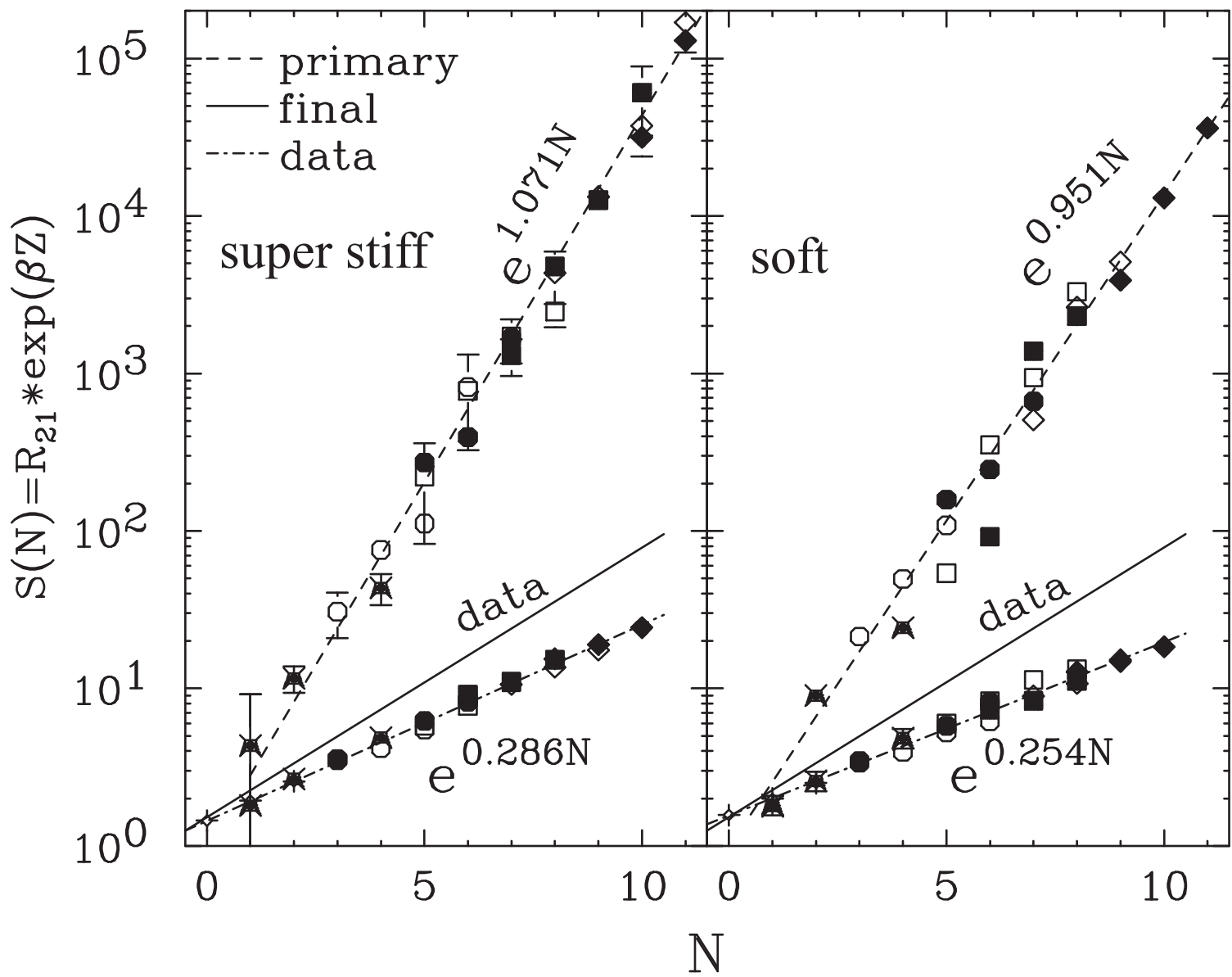
\end{document}